\documentstyle[11pt,aaspp4,psfig]{article}   

\slugcomment{Submitted to ApJ, 1999, September 27th}

\lefthead{Dieters et al.}
\righthead{4U\,1630-47: 1998 Outburst Timing Evolution}

\begin{document}


\def\about{$\sim$}
\def\rem#1{{\bf #1}}
\newcommand{\mdot}{$\dot{M}$}
\newcommand{\etal}{{\rm {et~al.}}}
\newcommand{\degree}{$^{\circ}$}
\newcommand{\rf}{\par\noindent\hangindent 15pt{}}

\title{The timing evolution of 4U\,1630-47 during its 1998 outburst}
\author{
        S.W. Dieters\altaffilmark{1,8}, 
        T. Belloni\altaffilmark{2}, 
        E. Kuulkers\altaffilmark{3}, 
        P. Woods\altaffilmark{1,8},
        W. Cui\altaffilmark{4}, 
        S.N. Zhang\altaffilmark{1}, 
        W. Chen\altaffilmark{5},
        M. van der Klis\altaffilmark{6}, 
        J. van Paradijs\altaffilmark{1,6}, 
        J. Swank\altaffilmark{5}, 
        W.H.G Lewin\altaffilmark{4},
	C. Kouveliotou\altaffilmark{7,8}
       }

\altaffiltext{1} {University of Alabama in Huntsville,
	301 Sparkman Dr., Huntsville, AL 35899, USA.}
\altaffiltext{2} {Osservatorio Astronomico di Brera, Via E. Bianchi 46,
	I-23807, Merate (LC), Italy}
\altaffiltext{3} {Space Research Organization Netherlands, Sorbonnelaan 2,
        3584 CA, Utrecht, The Netherlands, \& Astronomical Institute, Utrecht University}
\altaffiltext{4} {Massachusetts Institute of Technology, Center for
	Space Research, Cambridge, MA 02139, USA}
\altaffiltext{5} {NASA/Goddard Space Flight Center, Code 661, Greenbelt,
	MD 20771, USA}
\altaffiltext{6} {Astronomical Institute ``Anton Pannekoek'', University
	of Amsterdam, Kruislaan 403, 1098 SJ Amsterdam, The Netherlands}
\altaffiltext{7} {Universities Space Research Association}
\altaffiltext{8} {NASA/Marshall Space Flight Center, SD-50, Huntsville, AL 35812}

\authoremail{stefan.dieters@msfc.nasa.gov}


\begin{abstract}

We report on the evolution of the timing of 4U\,1630-47 during its 1998 outburst
using data obtained with the Rossi X-ray Timing Explorer (RXTE). The count rate
and position in hardness-intensity, color-color diagrams and simple spectral
fits are used to track the concurrent spectral changes. The source showed  seven
distinct types of timing  behavior, most of which show differences with the
canonical black hole spectral/timing states. In marked contrast to previous
outbursts, we find quasi-periodic oscillation (QPO) signals during nearly all
stages of the  outburst with frequencies between 0.06\,Hz and 14\,Hz and a
remarkable variety of other characteristics.  In particular we find large (up to
23\%\,rms) amplitude QPO on the early rise. Later, slow 0.1\,Hz semi- regular
short ($\sim$ 5 sec), 9 to 16\% deep dips dominate the light curve. At this time
there are two QPOs, one stable near 13.5\,Hz and the other whose frequency drops
from 6--8\,Hz to $\sim$ 4.5\,Hz during the dips. BeppoSAX observations during
the very late declining phase show 4U\,1630-47 in a low state.

\end{abstract}


\section {INTRODUCTION}

4U\,1630-47 has shown recurrent X-ray outbursts with intervals of 600--700\,days
(Jones et~al.\,1976, Priedhorsky 1986, Kuulkers et~al. 1997a). This is
much shorter than the typical waiting time of 10--50\,yrs for soft X-ray
transients (Tanaka \& Lewin 1995), which makes 4U\,1630-47 an important source
for studying how the spectral/timing behavior depends upon mass accretion rate
and/or other parameters.

4U\,1630-47 is a black-hole candidate (Parmar, Stella \& White 1986,  Barret,
McClintock \& Grindlay 1996, Kuulkers et~al.\,1997a) because it has shown 
spectral and timing characteristics similar to those of other X-ray sources 
with a measured high mass function.  The X-ray emission of black hole
candidates (BHCs) is usually classified into four timing/spectral states,
which are in order of increasing X-ray flux and presumably mass accretion rate:
low, intermediate, high and very high states (see van der Klis 1995; Cui
1999).  In the Low State (LS) the power density spectrum (PDS) shows strong
(30--50\%\,rms) band-limited (flat-topped) noise with a break between 0.03 and
0.3\,Hz and the energy spectrum is well fit with a power law with index of
$\Gamma=$$-1.5$\,--\,$-2.5$ up to at least 200\,keV (Gilfanov et~al.\,1994). In the High
State (HS) the PDS shows weak (few \%\,rms) power law noise. The energy
spectrum is dominated by a soft component which can be satisfactorily fit by a
multi-temperature disk blackbody.  The fitted temperature (kT$\sim$1\,keV) and
innermost radius are  comparable to values expected for a stellar mass black
hole. In the very high state (VHS) the soft component increases in flux  by a
further factor 2--8 over that in the high state. The power law component
strengthens and is visible in the energy spectra.  The power density spectrum
shows a variable broad band component and 3--10\,Hz QPO. The noise component is
either band-limited with a break between 1 and 20\,Hz, or a power law similar
to that in the  high state.  Occasionally at times between a HS and a LS an
intermediate state (IS) is observed e.g. GX\,339-4 (M\'endez \& van der Klis
1997), GS\,1124-68 (Belloni et~al.\,1997a), GRO\,J1655-40 (M\'endez, Belloni \&
van der Klis 1998) and Cyg\,X-1  (Belloni et~al.\,1996, Cui et~al.\,1997). The
intermediate state has fluxes and spectral properties intermediate those of the
HS and LS (Rutledge et~al.\,1999). The power spectral properties are very much
like the VHS. Slow $<$1\,Hz QPO have been observed in all states.

4U\,1630-47 has quite variable outbursts.  The 1996 outburst, lasted
$\sim$50\,days, the 1977 outburst lasted $\sim$10\,months and the 1988--91
interval showed ongoing long term activity (Kuulkers et~al.\,1997b). Even though
the 1987 and 1996 outbursts had similar flux histories  they show opposite
trends in spectral hardness as a function of time (Kuulkers et~al.\,1997a). Such
a range in behavior is akin to the variety of outbursts of the superluminal jet
sources GRS\,1915+105 and GRO\,1655-40. 

There are other similarities between 4U\,1630-47 and these sources. During the
1998 outburst radio emission was detected just as the source was making a
transition from hard to soft emission (Hjellming et~al.\,1999). This is very
similar to the behavior of GRS\,1739-278 and jet sources. Also 4U\,1630-47
showed strong  linear polarization in its radio flux, which has only been
observed  in elongated jet sources like SS433, GRS\,1915+105 and GRO\,1655-40
(Hjellming et~al.\,1999 and references therein).

Absorption dips lasting 50--150\,sec and reaching 8--30\% of the non-dip flux
have been observed from 4U\,1630-47 (Kuulkers et~al.\,1997b, Tomsick, Lapshov \&
Kaaret 1998). These dips, indicate that like GRO\,1655-40 and the accretion disk
corona  sources, 4U\,1630-47 is being viewed at a high (60--75\degree)
inclination. This cannot be confirmed since no optical counterpart has so far
been identified (Parmar et~al.\,1986, Buxton, private communication). Also it is
expected that these absorption dips will occur over a restricted range in phase
and so yield an estimate of the orbital period. We find no  evidence for any
absorption dips in our data.

The 1998 outburst began with a slow rise in the 20--100\,keV (BATSE) flux starting
near January 28 (TJD 10841; TJD=JD-2,440,000.5). Later on February 3 (TJD
10847) the RXTE all sky monitor (ASM; Levine et~al.\,1996) began showing
2--12\,keV flux. Radio emission is thought to have started just as soft flux began
its rapid rise near February 7 (TJD 10851). RXTE PCA observations were triggered at
this time. This outburst was well covered (100 observations) by RXTE (this paper,
Tomsick \& Kaaret 1999), ROSAT HRI/PSPC (between February 17 and 28), BeppoSAX 
(Oosterbroek et~al.\,1998), ASCA (February 25--26) and BATSE and radio (Hjellming
et~al.\,1999). In this paper we discuss the evolution of the timing behavior
during the 1998 outburst. 

Between May and September 1999, well in advance of its next expected outburst
(mid-December 1999; Kuulkers et~al.\,1998) 4U\,1630-47 underwent a fainter than
normal outburst. In terms of the very early flux evolution in the BATSE and RXTE
ASM energy ranges, the 1999 outburst was almost identical to that in 1998. On the
very early rise (1999 May 8) a single 0.85\,Hz, 16\%\,rms  QPO peak was found in
the power density spectrum (McCollough et~al.\,1999). At this time the energy
spectrum was hard. An observation two days later failed to find any QPO. 

\section{OBSERVATIONS}

The observations discussed here  (Obs. ID: 30178-0[1-2]- and 30188-02-) were
taken with the RXTE PCA (Jahoda et~al.\,1996) roughly daily, beginning on 1998
February 9, till 4U\,1630-47 reached maximum on February 22. After maximum
our observations were made every  2--5 days till 1998 March 31.  Observations
that partially overlap ours and extend until 1998 June 8 are reported by
Tomsick \& Kaaret (1999).

We analyzed 49 observations, most of which are $\sim$1.4\,ksec in length.  For
each of the pointings we calculated power density spectra from segments of data 
1,\,2,\,16,\,256 seconds long, using the whole energy range 2--60\,keV. The time
resolution is mostly 2\,ms. Power spectra were selected by time and intensity so
that the resulting average spectra are representative of the different flux
levels within each observation. In many observations the count rate did not vary
much and one average power density spectrum was sufficient to characterize the
observation.  The averaged power density spectra were normalized to the rms
variability (Belloni \& Hasinger 1990) using the average background  as estimated
using the PCABACKEST V2.1b, and corrected for dead-time (Zhang et~al.\,1995,
Zhang 1995). 

The average count rate, two hardness ratio's, and PHA spectra  were determined
for each observation using Standard\,2 mode data from  only a single PCA detector
(PCU\,0). Data selection and background subtraction followed standard
procedures.  

In addition we present two serendipitous BeppoSAX observations made with the 
MECS (Boella et~al.\,1997) on 1998 August 7 (TJD 11032) and 1998 September 16
(TJD 11072).  SAX was pointed at SGR\,1621-47, placing 4U\,1630-47 at the edge
of the field of view with an offset of 22$'$. The on source times were 85.7 and
62.4\,ksec, respectively. Source counts were collected from a 30$'$ diameter
circle that included the entire elongated image of 4U\,1630-47. Background
counts were extracted from a circular patch of the same size centered at the
same offset on the opposite side of the MECS field of view. The response matrix
was that for a 20$'$ offset. The expected resulting systematic errors on the
spectrum are 5--10\%. 

\section {RESULTS} 

During the outburst, seven distinct types of timing behavior occurred. We label
the different behaviors as A,B(b),C,D,E,F,G. The distinctions are primarily in
the shape of the broad band noise, and secondarily in the properties of the
QPO. The timing properties change hand-in-hand with changes in the energy
spectrum. Figure 1 gives an overview of the count rate and spectral changes as
represented by two lightcurves, a hardness-intensity and a color-color diagram.
Figure 2 shows from our data, five representative power density spectra
(behaviors A,B,C,D,E), whose fit parameters are given in Table 1.

\subsection{Timing properties}

{\bf A} During the first RXTE observation (1998 February 9, TJD 10854) while 
4U\,1630-47 was in the initial stages of its rise to maximum, a pair of very
strong (25\% and 5.6\% rms) QPOs were discovered in the power density spectrum
(Dieters et~al.\,1998a) at frequencies of 2.67 and 5.62\,Hz respectively. 
These are the first QPOs found from this source (except the marginal detection
of Kuulkers et~al.\,1998) and they are amongst the largest amplitude QPOs ever
measured from a BHC  (see Fig.\,2, panel A). Similar large amplitude QPO were
observed in the second observation. The power density spectra were fit with 3
Lorentzians to describe the peaks and  a broken (flat topped) power law to
describe the broad-band noise.  The lower frequency peak was asymmetric with a
shoulder toward higher frequencies. A Lorentzian was used to fit this shoulder.
The central frequencies of the two QPOs are not formally consistent with the
peaks being harmonically related but the frequencies are commensurate to well
within the widths of the QPO peaks.  The quality factor Q (ratio of frequency
to FWHM), is 12 and 8 in the 1st and 2nd observations for the lower frequency
QPOs and 3--5 for the shoulder and upper frequency QPO. Between the first two
observations the count rate increased by a factor of 1.26, the QPO frequencies
increased  (lower QPO: 2.67 to 3.20\,Hz, upper QPO: 5.62 to 6.66\,Hz), the
amplitude of the QPOs decreased (lower QPO: 17.03 to 15.29\%\,rms, upper QPO
5.79 to 4.16\%\,rms), the amplitude of the broad band noise remained roughly
constant at 20\%\,rms, and the break frequency  increased from
0.67$\pm$0.03\,Hz to 0.94$\pm$0.06\,Hz.

{\bf B} After the first two observations, there was a 1 day gap in RXTE
observations, during which 4U\,1630-47 brightened rapidly (Fig.\,1). Between
TJD 10856 and TJD 10864, 4U\,1640-47 continued to brighten, but at an ever
decreasing rate, forming a ``shoulder'' in the overall light curve. Short,
sharp dips appeared. These dips were generally triangular in shape with a sharp
drop ($\le$1\,sec) followed by a rise lasting 3--5\,sec. Initially they were
weak and fairly sporadic, but as the count rate increased they became more
regular, more closely spaced and deeper (9--16\%), dominating the light curve
(Fig.\,3). Toward the end of the shoulder the depth and the frequency of these
dips started to decrease.  In the power density spectrum these dips show up as
a QPO peak near 0.1\,Hz and occasionally its harmonic (See Fig.\,2, panel B).
The underlying power density spectrum is again modeled as a broken  power law
but the power density spectrum below the break (generally at 0.1--0.8\,Hz) is
not flat (index mostly 0.1--0.5). The noise amplitude is much weaker
(4.6--9.7\%\,rms) than in the initial (A) observations.  Three peaks are
observed in the power density spectra, typically at: 3--5\,Hz  (increasing with
count rate), 6--8\,Hz (no long term trend with count rate), and near 13.5\,Hz
(roughly constant), with rms amplitudes of 7--4\%, 4--1\%, $\sim$2\%,
respectively. The amplitudes generally decreased with count rate and time. The
QPO near 13.5\,Hz is weak and can only be detected in our longer observations.
Because of the gap in observations, we cannot tell if these three QPO are the
same as those seen in A. The highest frequency QPO has Q $\sim$9.  The Q of the
lower two peaks (2--7) is lower than the QPOs of A. This is especially true for
the lowest frequency peak.

We find two observations (denoted as b in Fig.\,1) outside the shoulder in
which dipping behavior reappears; one near maximum (C) and the other on the
decline (D). In both cases the count rates  and colors (spectra) become
comparable to those in the plateau. Within the Tomsick \& Kaaret data set there
are two further instances where dipping reappears on the decline. These
observations were made on the two days preceding our observation. 

We investigated the behavior of the QPOs in and outside the dips using 1 and 2
second data segments sorted by count rate.  We found that the $\sim$13.5\,Hz QPO 
remained constant in amplitude and frequency. However, the QPO peak with a
frequency near 6--8\,Hz, dropped in frequency to $\sim$4.5\,Hz within the dips
while remaining at a similar amplitude. There are only 2 QPOs: one stable near
13.5\,Hz, the other moving from 6--8\,Hz outside the dips to 4--5\,Hz. This
explains the 3 peaks in Fig.\,2B at 13.6,\,7, and 5\,Hz.  The width of the QPO
peaks is larger than just that expected from the QPO being truncated on entering
and leaving the dips.  This is consistent with either the frequency varying, or
the wave-trains that make up the lower frequency QPO being in general shorter
than the dips, i.e. $<$5 seconds. 

{\bf C} On about 1998 February 18, (TJD 10863, Fig\,1), the ASM/PCA count
rate increases rapidly reaching a broad roughly constant maximum until February
24 (TJD 10869), after which  the count rate drops rapidly. Here there  are no
dips and the power density spectrum shows weak (0.5--2.3\%), broad (few Hz) QPO
with frequencies in the 6.5--9.5\,Hz range superimposed upon a power law (slope
1.0--1.9) continuum (1.4--2.7\%\,rms.), see Fig.\,2C. Occasionally, an extra
($<$\,2\%\,rms) Lorentzian component peaking in the 1--2\,Hz was required in the
fits. This peak is similar to the marginal QPO detection of Kuulkers
et~al.\,(1998).

{\bf D} After the sharp drop ($\le$1\,day) on TJD\,10869, the flux continues to
decline  with count rates ranging from those of B to A. The power density
spectrum shows a broad bump near 1\,Hz, there is a steep component at very low
($<$0.1\,Hz) frequencies and a 2\%\,rms amplitude QPO peak near 11\,Hz. Note
that this steep component is similar in amplitude and slope to the power law
component of C.  The broad band noise was modeled with either a flat-topped
power law with a break near 1\,Hz and a zero-frequency Lorentzian or by a power
law and a broad Lorentzian peak near 1\,Hz (see Fig.\,2D). The broad bump
contributed 2.4--4.4\%\,rms, generally increasing as flux declined.  The very
low frequency component became generally weaker (max 1.8\%\,rms.) as
4U\,1630-47 faded. This sort of power density spectrum persisted even as the
count rate dropped to levels where large amplitude QPO (A) were observed on the
rise.  The colors in these decline observations are different from those of the
same flux on the rise. 

{\bf E} In the last two observations beyond TJD 10850  the very low frequency
component disappears and the power density spectrum can be well fit by just a
flat-topped broken power law.  The change from D to E seems to be gradual and
continuous. The break frequency is near 1\,Hz, the slope above the break
$\sim$1.5, and the amplitude $\sim$3\%\,rms. This flat-topped noise is very
similar to the $\sim$1\,Hz bump seen on the decline (D) and could well be the
same component. The absorbed flux (2--10\,keV) in our last observation (TJD
10873) is $1.9\times10^{-9}$erg\,cm$^{-2}$\,s$^{-1}$. 

{\bf F} Much later (at TJD 10950), Tomsick \& Kaaret (1999) report a marked
change in the power density spectra; from spectra with $\le$4\%\,rms noise
(like E) to spectra with $>$10\%\,rms broad band noise and a single 3.38\,Hz,
QPO with 19\%\,rms amplitude. The QPO frequency  decreased to about 0.2\,Hz as
4U\,1630-47 faded to a absorbed flux (2--10\,keV) of $\sim3\times10^{-10}$
erg\,cm$^{-2}$\,s$^{-1}$ at TJD 10972  (their last observation, and last
scheduled for RXTE). 

Eighteen days later (1998 June 26, TJD 10990) during an RXTE PCA observation of 
SGR\,1627-41 which included 4U\,1630-47 within the field of view (FOV; response
37\%) 0.15\,Hz QPO were found (Dieters et~al.\,1998b). The amplitude was at
least 3\%\,rms. The actual amplitude is dependent upon the relative
contributions to the flux from SGR\,1627-41, 4U\,1630-47 and the local galactic
ridge emission. Given that  BeppoSAX observations made on the 1998 August 7
(TJD 11032) and 1998 September 16 (TJD 11072) show that 4U\,1630-47 was still
much brighter than SGR\,1627-41, and that SGR\,1627-41 was in quiescence and
hence fainter (Woods et~al.\,1999) while Tomsick \& Kaaret (1999) found
3.4--0.2\,Hz QPO, it is reasonable to suppose that the 0.15\,Hz QPO originate
from 4U\,1630-47. Thus the amplitude of the 0.15\,Hz QPO must be at least
8.2\%\,rms.  

\subsection{Spectral evolution}

For each observation, the energy spectrum is modeled with the combination
of a multi-color disk component and a power-law component, which is typical for
black hole candidates  (Tanaka \& Lewin 1995). Such parameterization can
adequately describe  the observed X-ray spectrum of 4U 1630-47, although an
additional Gaussian  component is always required, probably indicating the
presence of an iron $K_{\alpha}$ line. Significant residuals still remain in
some  cases (especially C and D), but the results are good enough to provide a
rough description of the spectral evolution of the source during the outburst.
More detailed spectral results and discussions will be presented elsewhere
(Cui, Zhang, Sun et~al.\,1999, in preparation). The decline is also partially
covered by BeppoSAX observations (Oosterbroek et~al.\,1998; their observations
1$\cdots$5 occur when we find 4U\,1630-47 in behaviors to C, C, D or possibly
b, D, E respectively).

Following the onset of the outburst (the first four observations,  covering A
and the transition into B), both the soft component and the  power-law
component strengthen. Initially (A) the soft component is the weaker,
contributing $\stackrel{<}{_{\sim}}$20\% of the total (1.5--60\,keV) flux. At
the same time, the power law  steepens (the photon index goes from $-2.0$ to
$-2.5$) as shown by the changes in the color-color diagram of Fig.\,1. This is
consistent with the results from simultaneous ASM and BATSE observations
(Hjellming et~al.\,1999).  During B, the two components both grow in strength
with each contributing about half the unabsorbed flux.

Moving to C, the absorption increases (9--10 as opposed to
8--8.5$\times10^{22}$ atoms cm$^{-2}$) the power law  component becomes stronger
($\sim$60\%) relative to the soft (disk) component. As C progresses, the power
law component weakens and becomes steeper (photon index reaches as steep as $-2.8$). The
combination of higher absorption and steepening power law move 4U\,1630-47 away
from the pure-power laws toward the pure blackbody curves on the color-color 
diagram (Fig. 1).

The decaying period (D and E), begins with the soft component contributing most
of the flux (60--70\%) at a somewhat lower temperature (kT=1.2 keV rather than
1.4--1.5 of C). As the count rate drops, the absorption also drops,  the soft
component weakens and becomes softer while the power law component flattens. 
In E the soft component contributes at most 45--55\% of the flux, the
absorption is (4--7$\times10^{22}$ atoms cm$^{-2}$) and the power law index is
$\sim-2$.  This combination of changes move 4U\,1630-47 to the right and lower
on the color-color diagram (Fig.\,1).  These trends continue in F, with the
soft component contributes at most 10\% of the 2--60\,keV flux, the power law
index being in the range $-1.4$ to $-1.8$ and the absorption a little lower
still. Thus the energy spectra indicate a further hardening of the source i.e.
more like A. 

\subsection{The BeppoSAX data}

{\bf G} The BeppoSax MECS(2\&3) PHA spectra can be adequately fit (August:
reduced $\chi^{2}=0.8757$ with 92\,dof, September: reduced $\chi^{2}=1.5168$ with
92\,dof)  with a simple absorbed power-law model.  We used 3 and 4 channel
re-binning over the 1.8 to 11.5 keV range. The best fit parameters for 1998
August 7 (TJD 11032) and 1998 September 16 (TJD 11072) are
N$_{H}=7.7\pm0.45\times10^{22}$ and $8.1\pm0.7\times10^{22}$atoms\,cm$^{-2}$,
$\Gamma=1.42\pm0.07$ and $1.6\pm0.25$, with absorbed fluxes (2--10\,keV) of
$1.7\times10^{-10}$ and $0.98\times10^{-10}$ erg\,cm$^{-2}$\,s$^{-1}$,
respectively. These fluxes indicate a steady and long term decline in the
brightness of 4U\,1630-47 (Tomsick \& Kaaret 1999). Using an assumed distance of
10\,kpc the latter flux corresponds to a luminosity of  $1.2\times10^{34}$
erg\,s$^{-1}$. For a direct comparison with Parmar et~al.\,(1997) the luminosity
in September over the 2--2.4\,keV range with an absorption of 
$0.2\times10^{22}$atoms\,cm$^{-2}$ was $6\times10^{34}$ erg\,sec$^{-1}$.  This is
much brighter than the quiescent levels found by Parmar et~al.\,(1997). 

We calculated power density spectra for each BeppoSAX observation for the source
and background. Of the two observations only the August observation had
significant ($>5\sigma$)  excess power. At this time there was at least 50\%\,rms
variability (4U\,1630-47 \& background) over the 0.004--128\,Hz range. Most of
the excess is below 5\,Hz. The background (7\% of the source count rate) could
contribute at most ($3\sigma$ upper limit) 10\%\,rms to the source variability.
Therefore, 4U\,1630-47 had at least 40\%\,rms variability in the 0.004--128\,Hz
range. The combination of  high rms variability, and a power law energy spectrum
indicate that 4U\,1630-47 entered the low state.

\section{DISCUSSION}

As compared to previous observations of 4U\,1630-47, what is surprising about
the 1998 outburst is the sheer variety of timing behavior.  We can tentatively
link some of the QPOs seen from 4U\,1630-47. The large amplitude QPO seen on
the early rise (A) is similar in behavior to the 0.1--10\,Hz QPO on the early
rise of XTE\,J1550-564 (Cui et~al.\,1999). In both cases hard X-ray flux as
seen by BATSE is  declining and the spectrum in the PCA is softening.  The
3.7--0.2\,Hz QPO seen late in the outburst (F) by Tomsick \& Kaaret (1999) and
a 0.15\,Hz QPO later by Dieters et~al.\,(1998b) would be the same but in
reverse.  Given the evolution of the spectrum (soft to hard), this is
reasonable. Finally, the large amplitude QPO seen on the early rise of the 1999
mini(?)-outburst (McCollough et~al.\,1999), would also be the similar type of QPO.
In all cases the shape and amplitude of the broad-band noise is similar.
However, the harmonic content of the decline QPO and those in 1999 are
different from those on the rise. The appearance of these large amplitude QPO
with strong flat-topped noise  can not depend solely on mass accretion rate as
measured by count rate, since they do not reappear during the decline D. These
QPOs could well only appear in a restricted range in the color-color diagram
and therefore of spectral parameters.  Unfortunately, we cannot link the QPO of
A with those of B as would be suggested by the analogy with XTE\,J1550-564
which showed QPO changing continuously from 0.1 to 10\,Hz.

The quasi-periodic dips of B and b appear only when 4U\,1630-47 is in a fairly
narrow range of count-rate, and hardness. Their increase and then decrease in
depth and frequency as the count rate and color-color diagram position steadily
changes in B, indicates that there is some optimum conditions for their
occurrence. The pair of QPOs, one stable at 13.5\,Hz, the other with a  count
rate/frequency  dependence, are associated with the same conditions, since they
reappear in the b observations which are B-like excursions from C or D where
neither QPO are present.

The 1998 outburst of 4U\,1630-47 was clearly very complex. According to the
canonical model for black-hole transients, as a function of increasing
accretion rate a source should pass through the states in this order:
LS/IS/HS/VHS, and of course in reverse order as accretion rate decreases again
towards quiescence. As a first approximation, we can infer accretion rate from
the observed X-ray flux, although this approach has to be followed carefully.
During the course of our observations, covering most of the rise and decline (A
through E), there is always a mixture of a soft and hard component.

At the beginning of the outburst (A), at a relatively low  accretion rate, the
colors indicate that the source is rather hard, with the energy spectrum
dominated by a power law component (see Fig 1d) i.e. the soft component is $<$
20\% of the flux. The power density spectrum is similar to that of an IS 
(Belloni et~al.\,1997b), although the break frequency of the broad band noise
component is rather low, possibly connected to the fact that these observations
are made early in the outburst, when the source was moving from a LS to an IS.

Rather rapidly, the source increased in flux and moved to observations of type
B. There is a transition period as both the relative contribution of the soft
flux increases and the dipping behavior becomes established. Through B the soft
flux increases due to a combination of steepening in the power-law  and an
increase in the flux from the soft spectral component.   But at no time does
the soft component dominate the flux.   The power density spectrum is extremely
complex and not easy to interpret. The dipping behavior is unlike the
absorption dip(s) reported during the 1996 outburst. Superficially the dipping
behavior appears like that of GRS\,1915+105 (Belloni et~al.\,1997b) and 
GRO\,1655-40 (Remillard et~al.\,1999) for which the evidence is that the dips
are due to a disk instability. However, note that another and different type of
dips are present for GRS\,1915+105 (Belloni et~al.\,2000). Alternatively,  the
dips of 4U\,1630-47 may be more like dips and ``flip-flops'' seen in the VHS
of  GX\,339-4 (Miyamoto et~al.\,1991). A detailed comparison between these
various dipping behaviors is needed. In addition, there are 3 peaks in the
power density spectrum: one from a stable QPO near 13.5\,Hz and the other two
from an intensity-dependent QPO. The difference in QPO behavior with count rate
suggests that the two QPO have different origins.  There are numerous examples
of QPO in this range, but only those  of GRO\,1655-40 show a combination of
stationary and moving QPO (Remillard et~al.\,1999). However, the energy spectrum
of GRO\,1655-40 at this time shows the soft component  dominating (like VHS)
and there were no dips.   The combination of dips and QPO initially suggests
that type B behavior is a VHS but unlike the canonical VHS the soft
component does not dominate the flux, and it is not the highest flux regime.
Overall these ``shoulder'' observations do not fit in a clear way any of the
canonical states and seem to be peculiar.

Next, the source jumps to C and in terms of the  broad band noise into an obvious
HS.  The power-law shape, its slope and amplitude are a good match  to high
states in both 4U\,1630-47 and other BHC. The QPO is so weak that it could well
have escaped detection by previous instruments.  They are similar to those seen
from GRS\,1915+105 in the ``soft branch'' (Chen et~al.\,1997). However, the
spectrum indicates a strong power law component contributing $\sim$60\% of the
unabsorbed flux which is unlike a HS.  The transitions in and out of C are both
$\le$1\,day, and amongst the most rapid ever seen.

The exit from C into D is marked as a drop in count rate, softening of the disk
(soft) component, and a lowering of the absorption. Initially the soft component
contributes most of the flux  (60--70\%), but this contribution declines as the
count rate drops.  Also the power law hardens.  The power density spectra (see
Fig. 2D) are not unlike some observed in GX\,339-4 in its VHS (Miyamoto
et~al.\,1991 i.e. their Fig 4b).  Since the fluxes are lower in D and follow a
probable HS, we can exclude a VHS and it is natural to identify these observations
with an IS, which is known to show characteristics very similar to those of the
VHS (van der Klis 1995). The changes in the spectrum (HS-like toward LS-like)
and color-color diagram  support the identification with an IS.

At later times (E), the power density spectra are not very detailed due to the
lower statistics, but sre similar to the IS spectra reported by Belloni
et~al.\,(1997a) and M\'endez \& van der Klis (1997). The colors and spectra
indicate that the soft component decreased in relative strength and softened
further as the power law flattened. The source seems to be on its way, via the
reappearance of strong QPO and strong broad band noise, to the LS (high rms, power law
energy spectrum), which was reached by the time of the BeppoSAX observations (G)
well below the detection threshold of the PCA but much brighter than the
quiescent emission (Parmar et~al.\,1997). 

On the rise, the spectral evolution, the QPOs and their frequency evolution  are
very similar to that of XTE\,J1550-564 (Cui et~al.\,1999). This may not be that
surprising given that both these sources have been caught very early in their
outburst. In the past the wealth of information found during the rise phases of
the outburst have been missed. The changes in spectrum (soft to hard) on the
decline are similar to that of many transients. However, the power density
spectra are unusual. They are similar to that of Cir\,X-1 as it moves from the
``upper banana'' (similar to HS) through the ``lower banana" into an  extreme
``island state'' (similar to LS) (Oosterbroek et~al.\,1995). This serves as a
warning that the spectral/timing properties discussed here  and for other sources
are not unique to BHCs. 

\acknowledgments

This work supported by NASA/LTSA-NAG5-6021(Dieters, van Paradijs, Kouveliotou \&
Lewin), NAG5-7483 (Dieters), NAG5-7484 (Cui)  and by the Netherlands
Organization for Scientific research (NWO) under grants 614-51-002 and
Spinoza-08-0.

\newpage

\figcaption[]{
The left-hand panels show the ASM and PCA light curves for the 1998 outburst. On
the ASM light curve (top left panel) the times of all pointed observations are
marked.  The PCA count rates (lower left) are the background subtracted rates,
over the full energy range ($\sim$2--$\sim$60\,keV), for PCU\,0 only. The two
right-hand panels show the hardness-intensity and color-color diagrams. The count
rates and hardness ratio's are from PCU\,0 only. The positions of the various
timing/spectral behaviors are labeled, as in the text {\bf A,B,C,D,E}. Label
''{\bf b}" indicates the reappearance of the dipping behavior of {\bf B}. In the
color-color diagram two sets of theoretical curves are shown: for a pure disk
black-body and for a pure power law. Two different absorptions are shown:
N$_{H}=8\times10^{22}$  (dotted) and N$_{H}=10\times10^{22}$atoms\,cm$^{-2}$
(dot-dashed). As a reference,  the rightmost points for the power-law model
correspond to a photon index of 2.2 for  the N$_{H}=8$ and 2.3 for N$_{H}=10$.
The other points  steepen in steps of 0.1 and go until $\gamma=4.0$. The disk
black body points close to  Y=1.0 are kT=1.4 (N$_{H}=8$) and kT=1.3 (N$_{H}=10$).
}

\figcaption[]{
Representative power density spectra in chronological order (top to bottom)
through the 1998 outburst. Throughout the text the different types of timing
behavior are referred to by their label within this figure i.e., A,B,C,D,E. }

\figcaption[]{
Light curve (Standard 1) over the 2$\sim$60\,keV range showing the dipping
behavior in B. The observation was made on 1998 February 13 (TJD 50857). }

\clearpage

\begin{deluxetable}{ccccccccc}
\footnotesize
\tablecaption{Broad band noise and QPO fit parameters.} 

\tablewidth{0pt}

\tablehead{}
\startdata

\hline
\multicolumn{6}{c}{Noise fit parameters} & 
\multicolumn{3}{c}{QPO fit parameters}\nl

                      &
Count rate            &
\% rms                &
$\nu_{break}$         &
$\alpha<\nu_{break}$  &
$\alpha>\nu_{break}$  &
\% rms                &
FWHM                  &
Frequency             \nl
                      &
c/s/5PCU's            &
0.01-100\,Hz          &
                      &
                      &
                      &
                      &
Hz                    &
Hz                    \nl

\hline

 A & 1426.34 & $19.01^{0.56}_{0.77}$ & $0.67\pm0.03$        & 0.0   Fixed           & $1.44^{0.03}_{0.07}$ & $17.03^{0.18}_{0.38}$ & $0.2113^{0.006}_{0.015}$ & $2.677^{0.0045}_{0.0054}$ \nl
   &         &                       &                      &                       &                      & $ 7.93^{0.66}_{0.83}$ & $0.95\pm0.013          $ & $3.33^{0.06}_{0.15}$	   \nl
   &         &                       &                      &                       &                      & $ 5.79^{0.41}_{0.48}$ & $1.49^{0.19}_{0.22}$     & $5.60\pm0.05$	   \nl
 B & 4710.85 & $2.09^{0.13}_{0.12}$ & $2.26^{0.10}_{0.14}$ & $0.51^{0.048}_{0.082}$ & $1.78^{0.19}_{0.14}$ & $2.876^{0.041}_{0.050}$  & $1.072^{0.045}_{0.045}$ &  $4.953^{0.017}_{0.017}$  \nl
   &         &                       &                      &                       &                      & $4.340^{0.032}_{0.070}$  & $3.321^{0.065}_{0.075}$ &  $7.048^{0.030}_{0.033}$  \nl
   &         &                       &                      &                       &                      & $2.035^{0.029}_{0.0485}$ & $1.682^{0.060}_{0.100}$ & $13.674^{0.022}_{0.018}$ \nl
   &         &                       &                      &                       &                      & $2.119^{0.071}_{0.0905}$ & $4.71^{0.85}_{0.50}$	&  $0.91^{0.18}_{0.10}$ \nl
 C & 5736.95 & $1.63^{0.15}_{0.13}$ &                      & $1.13^{0.09}_{0.08}$   &                      & $2.29^{0.25}_{0.19}$ & $6.9^{2.1}_{1.5}$    &  $8.16^{0.45}_{0.50}$ \nl
 D & 3724.48 & $1.35^{0.35}_{0.22}$ &                      & $1.35^{0.27}_{0.20}$   &                      & $2.28^{0.27}_{0.22}$ & $2.92^{0.46}_{0.42}$ &  $0.80^{0.20}_{0.28}$ \nl
   &         &                      &                      &                       &                       & $1.15^{0.12}_{0.10}$ & $5.26^{1.66}_{1.21}$ &  $10.54^{0.52}_{0.50}$ \nl
   & 3724.48 & $1.02^{0.09}_{0.07}$ & $1.39^{0.13}_{0.13}$ & 0.0   Fixed           & $1.585^{0.25}_{0.23}$ & $1.80^{0.15}_{0.15}$ & $0.0539^{0.0178}_{0.0168}$ & 0.0 Fixed \nl
   &         &                      &                      &                       &                       & $2.11^{0.32}_{0.25}$ & $4.8^{1.7}_{1.2}$	       & $10.59^{0.53}_{0.51}$ \nl
 E &  893.21 & $1.42^{0.22}_{0.22}$ & $1.28^{0.78}_{0.24}$ & 0.0   Fixed           & $1.38^{1.915}_{0.28}$ &			  &                            & \nl
\hline

\enddata

\tablecomments{
The broad band noise and QPO fit parameters of the 5 representative power
density spectra shown in Fig.\,2. A broken power law is used for A,B,E and a
power law is used for C to model the broad band noise. Lorentzian peaks are
used to model any QPO peaks and broader ``bumps". For A, the addition of a 0.34
FWHM bump at 0.55\,Hz with 8.8\%\,rms amplitude can also be added, but for only
the first observation. For D two equivalent models were used:  (top) a power
law with a ``bump" and a QPO, (bottom) a broken power law with a zero frequency
Lorentzian and QPO. The ``bump" and zero frequency Lorentzian can be considered
extra components of the broad band noise. The errors are the 1\,$\sigma$ single
parameter error i.e.; the parameter range within $\Delta\chi^{2}=1.0$ of the
best fit minimum.
 }

\end{deluxetable}

\newpage
\clearpage

\centering
\hbox{{\psfig{figure=f1a.ps,height=4cm,width=6cm,angle=-90}}
      {\psfig{figure=f1c.ps,height=4cm,width=6cm,angle=-90}}}
\hbox{{\psfig{figure=f1b.ps,height=4cm,width=6cm,angle=-90}}
      {\psfig{figure=f1d.ps,height=4cm,width=6cm,angle=-90}}}
\vskip 2 true cm
{Figure 1}

\newpage
\par
\centering
\hbox{{\psfig{figure=f2a.ps,height=4cm,width=4cm}}}
\hbox{{\psfig{figure=f2b.ps,height=4cm,width=4cm}}}
\hbox{\hspace{-0.5cm}{\psfig{figure=f2c.ps,height=4cm,width=4.5cm}}}
\hbox{{\psfig{figure=f2d.ps,height=4cm,width=4cm}}}
\hbox{{\psfig{figure=f2e.ps,height=4cm,width=4cm}}}
{Figure 2}

\newpage
\clearpage
\par


\centering
\hbox{{\psfig{figure=f3.ps,height=18cm,width=14cm}}}
\vskip 2 true cm
{Figure 3}

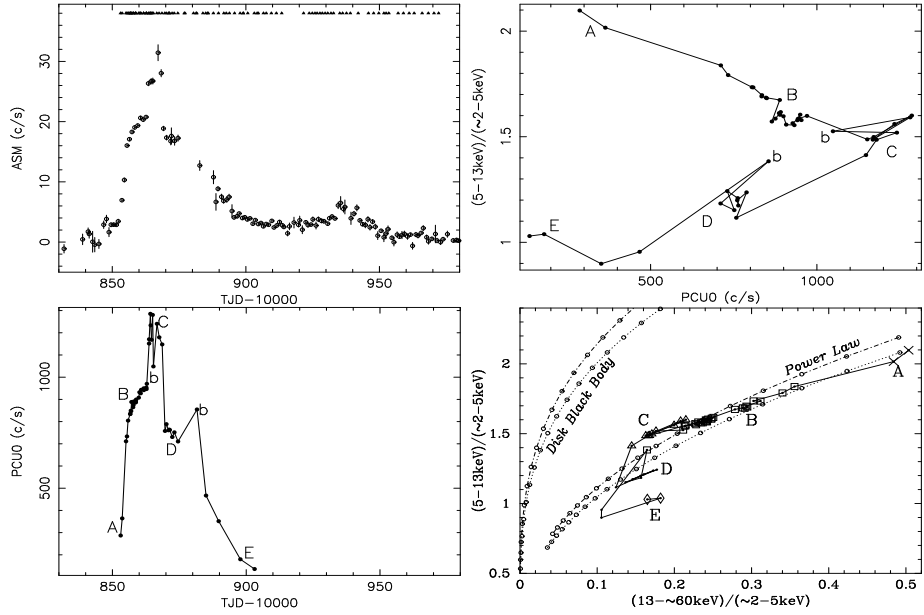
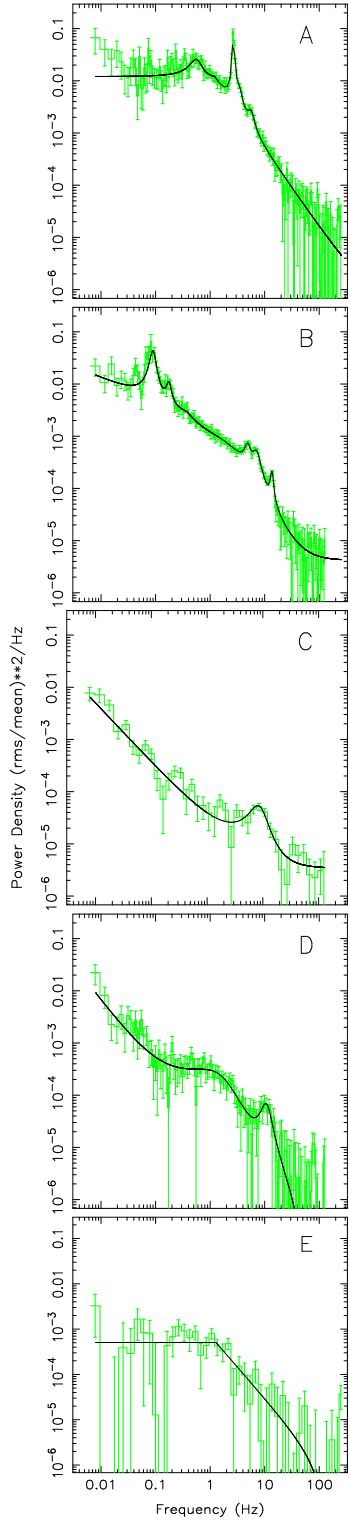
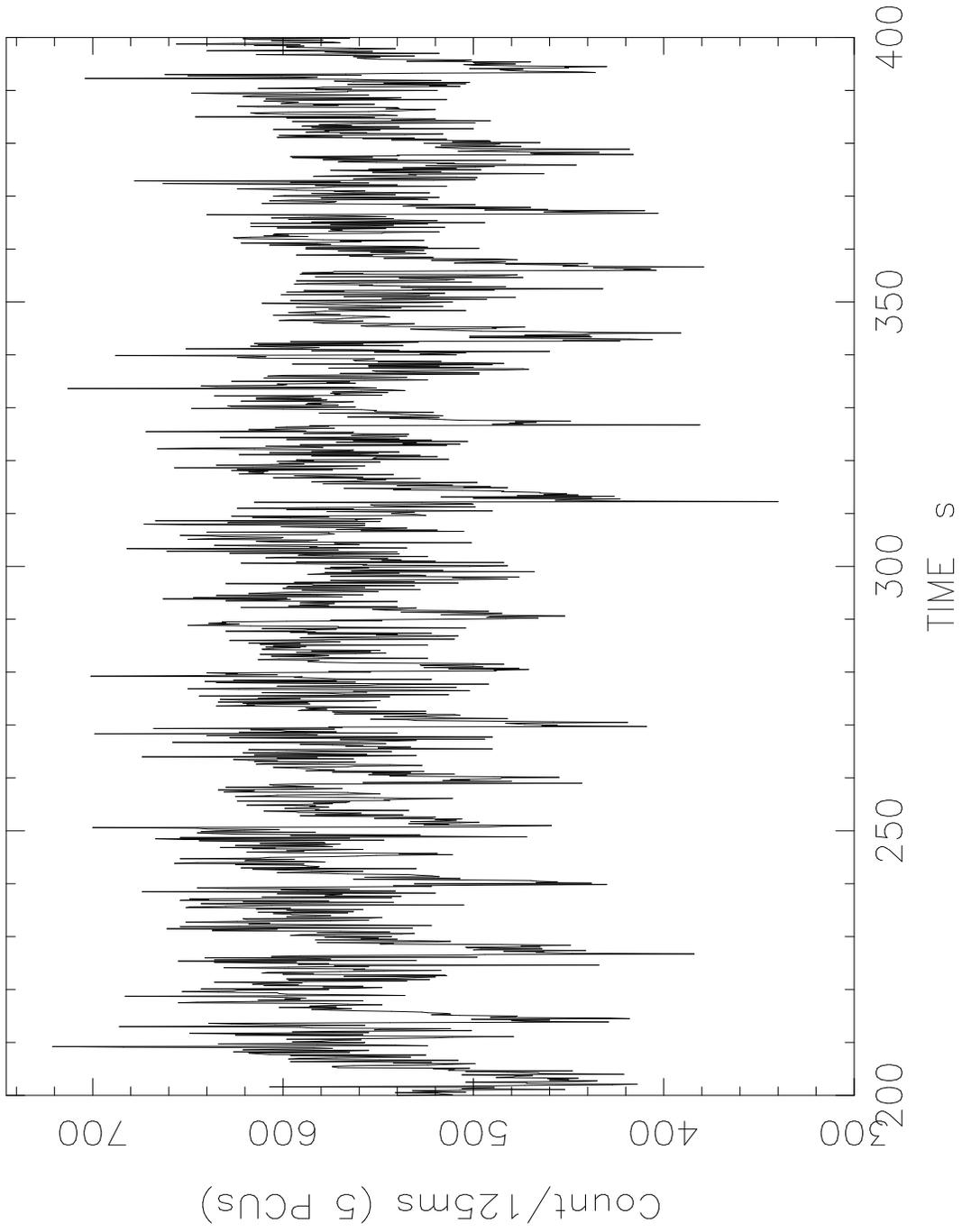
\end{document}